\documentclass{article}

\usepackage{psfrag}
\usepackage{amsmath}
\usepackage{amssymb}
\usepackage{wasysym}
\usepackage{dsfont}
\usepackage{url}
\usepackage{framed}
\usepackage[dvips]{graphicx}
\usepackage{rotating}
\usepackage{prettyref}

\usepackage{flexisym}
\usepackage{breqn}

\def\newblock{\hskip .11em plus .33em minus .07em}

\newcommand{\abs}[1]{\left|#1\right|}
\newcommand{\set}[1]{\left\{#1\right\}}
\newcommand{\To}{\rightarrow}

\newcommand{\E}{\mathds{E}}

\newcommand{\Prob}[1]{\Pr\set{#1}}

\newcommand{\ghgf}[5]{{_#1}F_{#2}\!\!\left(\!\!\!\!\begin{array}{c|c}
                                        \begin{array}{c}
                                          #3 \\
                                          #4
                                        \end{array}
                                         & #5
                                      \end{array}\!\!\!\!\right)}
\newcommand{\hgf}[3]{\ghgf{2}{1}{#1}{#2}{#3}}
\newcounter{thm}
\newtheorem{assumption}[thm]{Assumption}
\newtheorem{defn}[thm]{Definition}
\begin{document}

\title{Simple Error Scattering Model for improved Information Reconciliation}

\author{Stefan~Rass\\
System Security Group, Alpen-Adria University of Klagenfurt, \\
Austria, email: stefan.rass@uni-klu.ac.at}

\date{August 2009}

\maketitle

\begin{abstract}
Implementations of quantum key distribution as available nowadays suffer from inefficiencies due to post processing of the raw key that severely cuts down the final secure key rate. We present a simple model for the error scattering across the raw key and derive "closed form" expressions for the probability of a parity check failure, or experiencing more than some fixed number of errors. Our results can serve for improvement for key establishment, as information reconciliation via interactive error correction and privacy amplification rests on mostly unproven assumptions. We support those hypotheses on statistical grounds.
\end{abstract}

\section{Introduction} Quantum key distribution is an emerging technology that promises information-theoretic security, without computational infeasibility assumptions. Taking the classical BB84 protocol as an example, this one comprises the following steps:

\begin{enumerate}
  \item Alice sends randomly polarized photons to Bob
  \item Bob measures in a random polarization basis
  \item Bob keeps track of successful measurements
  \item Bob tells Alice the polarizer settings he used
  \item Alice tells Bob which settings have been correct
  \item Both discard all incorrect measurements
  \item Both map measurement results to a bitstring
  \item Both perform error correction
  \item Both perform privacy amplification
\end{enumerate}

In a nutshell, the idea is to exploit photons as carriers of information due to their incapability of being copied. This renowned result is known as the no-cloning theorem \cite{Wootters1982}. Hence, any unauthorized access to the information encoded in the photons will result in an unnaturally high error rate, thus revealing the eavesdropping attempt eventually. The original protocol is found in \cite{Bennett1984}. A rigorous proof of security is provided in \cite{Shor2000}, for instance.

We are particularly interested in step 8 of the BB84 protocol, in
which Alice and Bob locate and repair errors in their bit-strings. The process is known as \emph{reconciliation}, and involves Alice
and Bob publicly exchanging parity bits in order to correct errors and distill identical keys. With each published parity bit, a piece of
information leaks out and becomes visible to the adversary, which is the reason why reconciliation is followed by
\emph{privacy amplification}. Basically, this is the application of a strongly universal hash function on the output, in order to create dependencies of the final bits on the bits that the
adversary did not get any information about, thus sufficiently decreasing the adversary's information. We shall not go into details
about the privacy amplification, and refer the reader to the literature on universal hashing
\cite{Stinson1992,Wegman1981a,Carter1981} as well as theoretical results about privacy amplification
\cite{Renner2005,Renner2005a}, and references therein for further information.

\section{Interactive Error Correction}

Let us pay closer attention to the error correction mechanism which has been proposed along with the experimental
implementation of BB84 \cite{Bennett1992a}. Errors in physical transmission media often exhibit burst structures, that
is, a sequence of consecutive errors is more likely to occur than sparse scattering. Consequently, a popular trick is
to permute the bits in the string prior to the error correction in order to chop down long bursts into small pieces.
Ideally, this leaves an almost uniform pattern of erroneous bits in the result. This is the first step in a protocol
which has become known under the name \emph{Cascade}.

After having agreed on a publicly known permutation of bits,
Alice and Bob take their shuffled strings and partition them into blocks of constant size $k$, such that a single block
is believed to contain no more than one error with high probability. The protocol was first introduced in
\cite{Bennett1992a}.

The problem of how errors are scattered across the raw key has been tackled on statistical grounds in \cite{Gilbert2000}. The authors of this work assume a binomial distribution of errors within the blocks, which is later approximated by a Poisson-distribution.

We shall take a different route here, considering the process that induces the errors to be Poissonian, as well as adapting the initial block-size using a decision-theoretic approach. The authors of \cite{Gilbert2000} do not provide a direct clue on how to choose an optimal block-size for partitioning. This is the gap we intend to close now. Before getting into details about how to cleverly choose the block-sizes, let us outline the remaining steps in the error correction process. This will highlight the room for improvement that an intelligent partitioning strategy can exploit.

\begin{figure}[t!]
  \centering
  \psfrag{a}[c][]{\scriptsize{Alice}}
  \psfrag{b}[c][]{\scriptsize{Bob}}
  \psfrag{e}[c][]{\scriptsize{\checkmark}}
  \psfrag{u}[c][]{\scriptsize{$!$}}
  \psfrag{p}[c][]{\scriptsize{parity comparison ok (\checkmark) or mismatch ($!$)}}
  \includegraphics[width=0.8\textwidth]{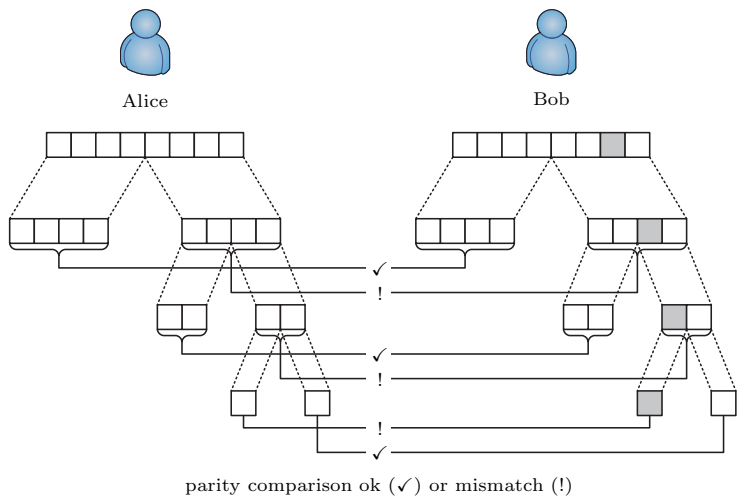}\\
  \caption{Bisective search for errors}\label{fig:bisective-error-search}
\end{figure}

Having split the string into blocks of equal size $k$, Alice and Bob publicly compare parity bits of each block.
Obviously, one error will change the parity, and in general any odd number of errors will be discovered by observing
disagreeing parities. However, two or any larger even number of errors will remain undetected with this method, which
is why further stages of the process are to follow, once the initial correction has been completed. For the correction
of errors, take an example-block with one indicated error, a block where a parity mismatch was observed during the public comparison. Then this block
is searched for the error using a standard bisective search, which discloses a further lot of $\log(k)$ parities of
sub-blocks. The process is depicted in \prettyref{fig:bisective-error-search}. To spot and remove remaining errors in the string, such as present in blocks with an even number of errors in
them, Alice and Bob repeat the shuffling and partitioning steps, several times with increasing block-sizes.

\textbf{Example}: The inherent problem with parity checking, which motivates the need for the repeated shuffling and
creation of larger blocks, can be illustrated as follows: assume that Alice and Bob share the following bit-strings,
with errors in Bob's string being underlined,
\[
  \begin{array}{rc}
    \text{Alice:} & 0011010100101101110101010011010\ldots \\
    \text{Bob:} & 001101\underline{1}100101101\underline{00}0101\underline{1}1\underline{1}0110\underline{0}0\ldots \\
  \end{array}
\]
Partitioning into blocks of equal size and comparing parities of blocks will let some errors go undisclosed (parity
bits are shown in the two middle rows; agreeing parities hiding existing errors are underlined):
\[
\begin{array}{rc}
  \text{Alice:} & \underbrace{00110}_0|\underbrace{10100}_0|\underbrace{10110}_1|\underbrace{11101}_0|\underbrace{01010}_0|\underbrace{01101}_0|0\ldots \\
  \text{Bob:} & \overbrace{00110}^0|\overbrace{1\underline{1}100}^1|\overbrace{10110}^1|\overbrace{1\underline{00}01}^{\underline{0}}|\overbrace{01\underline{1}1\underline{1}}^{\underline{0}}|\overbrace{0110\underline{0}}^1|1\ldots
\end{array}
\]
Sparing the shuffling of bits will save some errors from discovery, since two blocks (in this example, the 4th and 5th)
with even number of errors can form a larger block with an even number of errors again. Therefore, permuting the bits
is inevitable to avoid such an undesirable coincidence.

Since the error correction up to now may be ineffective, as still having missed some errors, Alice and Bob continue by
comparing parities of random subsets of bits they publicly agree on. Upon parity mismatch, a bisective search similarly
as above is performed to find and erase the error. In order to avoid information leaking to the adversary, the last bit
from each random subset is deleted. This deletion is also done after comparing parities of blocks in the previous steps, for the same reason.

The point at which almost all errors have been removed is detected by counting the number of successful comparisons
after having discovered the last error. After a sufficient number of successful trials (20 is the number proposed in
\cite{Bennett1992a}), the strings are accepted as identical, regarding the probability of errors remaining undetected
as negligible.

The protocol \emph{Cascade} is based on this procedure and has been introduced in a later paper \cite{Brassard1993},
which presented improvements to the original reconciliation protocol sketched above. Among the changes is the removal
of the bit-deletion step for the sake of detecting more errors faster, so the task of information leakage reduction is
shifted to the privacy amplification stage. The naming stems from the strategy of increasing sizes of blocks in the
first stages of the protocol. Although a theoretical analysis of the protocol is provided, the authors of
\cite{Brassard1993}, as well as those of \cite{Bennett1992a} abstain from an analytical treatment of block-size
choices. Nevertheless, they give a simple heuristic based on estimating the error frequency by direct comparison of a
random sample of bits is provided in \cite{Bennett1992a}. These bits have to be sacrificed for the sake of privacy too,
if that approach is adopted. To summarize, the error correction protocol in charge of current QKD implementations relies on

\begin{assumption}\label{ass:block-size} A block-size exists, such that by partitioning the raw key into blocks of that given size, each block contains at most one error.
\end{assumption}

It is this assumption that we seek to support by our upcoming theoretical considerations.

\section{An error scattering model}

We choose the Poisson process as the natural model for errors that occur within a sequence of bits that can be arbitrarily long.
\begin{defn}[Poisson process \cite{Ross1983}] A \emph{Poisson process} is a family of discrete counting measures
$\set{N(t):t\geq 0}$, which satisfy the following conditions:
\begin{enumerate}
  \item $N(0)=0$ (no events yet at the beginning).
  \item The process has independent increments.
  \item The number of events in any interval of length $\Delta t$ is Poisson distributed with mean $\lambda \Delta
      t$. That is, for all $t,\Delta t\geq 0$,
  \[
    \Prob{N(t+\Delta t)-N(t)=n}=\frac{(\lambda\cdot\Delta t)^n}{n!}e^{-\lambda\cdot\Delta t}, \quad n=0,1,2,\ldots.
  \]
  In other words, the expected number of events linearly increases with the length of the block.
\end{enumerate}
\end{defn}
Our Poisson process model $N(t)$ will count the total number of errors in the bit string at time $t$. Since our blocks
should be chosen such that the expected additional number of errors after taking a time-step $\Delta t$ is only $1$. The intensity parameter $\lambda$ determines the frequency of events, i.e. errors in our case.

\begin{figure}\centering
    \psfrag{r}[b][]{Intensity $\Lambda\sim\mathcal{G}(10,2)$}
    \psfrag{x}[c][]{Time}
    \psfrag{T}[t][]{\scriptsize{$T$}}
    \psfrag{0}[c][]{\scriptsize{$0$}}
    \psfrag{0.1}[c][]{\scriptsize{$0.1$}\hspace{2mm}}
    \psfrag{0.2}[c][]{\scriptsize{$0.2$}\hspace{2mm}}
    \psfrag{0.3}[c][]{\scriptsize{$0.3$}\hspace{2mm}}
    \psfrag{0.4}[c][]{\scriptsize{$0.4$}\hspace{2mm}}
    \psfrag{0.5}[c][]{\scriptsize{$0.5$}\hspace{2mm}}
    \psfrag{0.6}[c][]{\scriptsize{$0.6$}\hspace{2mm}}
    \psfrag{0.1}[c][]{\scriptsize{$0.1$}\hspace{2mm}}
  \includegraphics[width=0.8\textwidth]{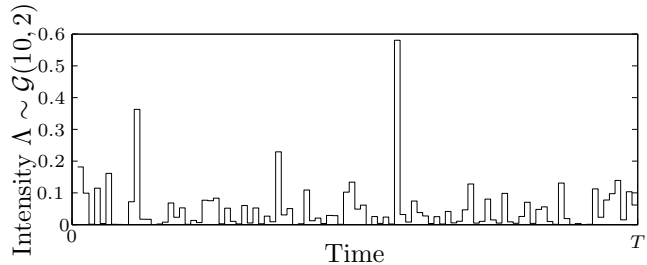}\\
  \caption{Example error intensity parameter process}\label{fig:rateprocess}
\end{figure}

Assume that this intensity-parameter remains constant over a unit of time, and denote it by
$\Lambda\in[0,\infty)$. \prettyref{fig:rateprocess} shows an example with finite time horizon $T$, and gamma-distributed error-rate (with parameters $a=10$ and $b=2$ for the gamma-distribution), remaining constant over short periods of time.

The probability of exactly $k$ errors within a time unit is found from the law of total probability: let $X$ be the
number of errors per time-unit, then
\begin{equation}\label{eqn:k-errors-per-block}
    \Prob{X=k} = \int_0^\infty \underbrace{\Prob{X=k|\lambda = x}}_{\text{\shortstack{Poisson distributed\\with parameter $\lambda$}}}\underbrace{\Prob{\lambda=x}}_{\text{\shortstack{intensity\\parameter}}}dx.
\end{equation}

\begin{figure}[htbp]
  \centering
    \psfrag{0.05}[c][]{\scriptsize{$0.05$}}
    \psfrag{0.15}[c][]{\scriptsize{$0.15$}}
    \psfrag{0.1}[c][]{\scriptsize{$0.1$}}
    \psfrag{0}[c][]{\scriptsize{$0$}}
    \psfrag{5}[c][]{\scriptsize{$5$}}
    \psfrag{15}[c][]{\scriptsize{$15$}}
    \psfrag{10}[c][]{\scriptsize{$10$}}
    \psfrag{20}[c][]{\scriptsize{$20$}}
    \psfrag{25}[c][]{\scriptsize{$25$}}
    \psfrag{y}[b][]{$\Prob{X=k}$}
    \psfrag{x}[t][]{$k$}
  \includegraphics[width=0.8\textwidth]{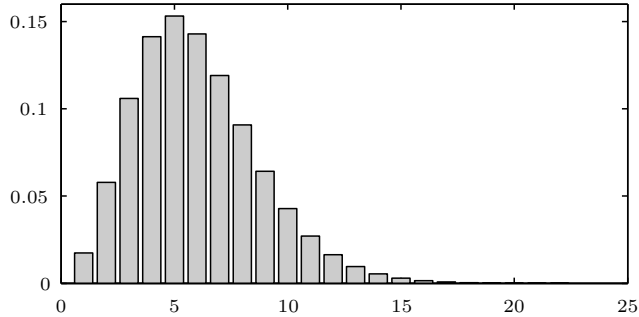}\\
  \caption{Example of error probabilities under Cox process error scattering}\label{fig:density-example}
\end{figure}

For the intensity-parameter, we assume a gamma-distribution. This choice is intuitively reasonable, as this class is flexible and supported on the nonnegative real line. Plugging into \eqref{eqn:k-errors-per-block} the density of
the Gamma-distribution given by
\[
f(\lambda|a,b) =
    \left\{
      \begin{array}{ll}
        \frac{b^a}{\Gamma(a)}\lambda^{a-1}e^{-b \lambda}, & \lambda\geq 0 \\
        0, & \lambda<0,
      \end{array}
    \right.
\]
and the density of the Poisson-distribution, which is
\[
    f(k|\lambda)=\frac{\lambda^k}{k!}e^{-\lambda},
\]
we find
\begin{eqnarray}
    \Prob{X=k} &=& \frac{b^a}{k!\Gamma(a)}\int_0^\infty x^{k+a-1}e^{-x(b+1)}dx\nonumber\\
    &=& \frac{b^a\Gamma(k+a)}{k!(b+1)^{k+a}\Gamma(a)}\label{eqn:error-probability},
\end{eqnarray}
where $\Gamma$ denotes Euler's Gamma-function. \prettyref{fig:density-example} shows an example of this density with
(arbitrarily chosen) parameters $a=10$ and $b=2$.

This discrete probability density has a closed form expression for its first moment. The expected number of errors per
time unit for this model is
\begin{equation}\label{eqn:first-moment}
    \E X=\sum_{k=0}^\infty k\cdot\Prob{\text{$k$ errors per time unit}} = \frac a b.
\end{equation}

So far, we are almost done, because Assumption \ref{ass:block-size} stated in the previous section can now easily be fulfilled: simply choose the block size inverse to the expected number of errors, which will eventually leave a single error per block. This can theoretically be justified by considering the following result, related to Poisson processes in general: as we explicitly know the expected number of errors within time-span $\Delta t$, which is
$\frac a b\cdot(\Delta t)$. Assuming that the bits come in at frequency $f$, then if $n$ denotes the number of arriving
bits within $\Delta t$, we have $f\cdot \Delta t = n$ and the block-size in terms of bits is the solution of the
equation $\frac a b \cdot\frac n f=1$, i.e.
\[
    \text{initial block-size }n\approx \frac f {a/b},
\]
which is the block-size (in bits) that the Poisson process gives us. Observe that we have a constant block-size again.
The only difference to the original Cascade variant is that it originates from a probabilistic model, rather than from
pure intuition.

All calculations above were done with the \textsc{Maple} software \cite{Maple10}. The density of $X$ can
be derived by appropriately substituting terms under the integral to obtain the same form as for a  Gamma-distribution (with different parameters, though). Then the normalizing constant takes the same form as for the
gamma-density, giving the result.

Using the density \eqref{eqn:error-probability}, we can give a formula for the probability of seeing more than $m$
errors during a time unit as
\begin{equation}\label{eqn:more-than-m-errors}
    \Prob{X>m} = \frac{b^a(a)_{m+1}}{(m+1)!(b+1)^{a+m+1}}\cdot\hgf{1,m+a+1}{m+2}{\frac 1 {b+1}},
\end{equation}
where $a,b>0$ describe the Gamma-distribution of the intensity-parameter, $_2F_1$ is the hypergeometric function (see
Equation \eqref{eqn:generalized-hypergeometric-function}), and $(a)_m$ is the Pochhammer symbol (see Equation
\eqref{eqn:pochhammer-symbol}).

Since error correction in the way used with quantum key distribution relies on public parity comparisons, the
event of missing an error is equal to the event of having an \emph{even} number of errors. The probability that the
parity check can indicate an error, is calculated as follows (the derivation is shown in Section \ref{sec:proofs}):
\begin{equation}\label{eqn:parity-miss}
    \Prob{\text{odd number of errors}} = p_{\text{odd}} = \frac 1 2\left[1-\left(\frac b{b+1}\right)^a\right].
\end{equation}

For a finite string of length $2m+1$, this probability is
\begin{equation}\label{eqn:finite-parity-miss}
p= p_{\text{odd}} - C(a,b,m)\cdot\ghgf 3 2{1,m+2+\frac a 2,m+\frac 3 2 +\frac a 2}{m+2,m+\frac 5 2}{\frac 1{(b+1)^2}},
\end{equation}
where
\[
C(a,b,m) = \frac{b^a\Gamma(2m+3+a)}{\Gamma(a)(2m+3)!(b+1)^{2m+3+a}}.
\]

The previous results are an appealing tool for a proper choice of the parameters if an error-correcting code shall be used with the scheme. Employing classical error correction
mechanisms may not work well, since our wish is to detect eavesdropping via a raised error rate. Therefore, we cannot
adopt any assumption on how many errors will occur at maximum, and classical error-correcting codes can no longer be
used for that matter. However, if the probability of seeing more than $m$ errors can be bounded, then such codes could become indeed applicable. We shall not go into further details about this here.

\section{Proofs}\label{sec:proofs}

This section is dedicated to proving equations \eqref{eqn:first-moment}, \eqref{eqn:more-than-m-errors}, \eqref{eqn:parity-miss} and \eqref{eqn:finite-parity-miss}.

For the probability distribution function let us first substitute $c:=b+1$ into \eqref{eqn:error-probability}, and set
\begin{eqnarray}
    F(m) = \Prob{X\leq m}=\sum_{k=0}^m \Prob{X=k} &=& \sum_{k=0}^m \frac{b^a\Gamma(k+a)}{k!(b+1)^{k+a}\Gamma(a)}\nonumber \\
     &=& \frac{b^a}{(b+1)^a\Gamma(a)}\sum_{k=0}^m\frac{\Gamma(k+a)}{k!(b+1)^k}\nonumber\\
     &\propto& \sum_{k=0}^m \frac{\Gamma(k+a)}{k!c^k}.\label{eqn:cox-distribution}
\end{eqnarray}
Using the \textsc{Maple} software \cite{Maple10}, we obtain for the finite sum \eqref{eqn:cox-distribution},
\begin{dmath}\label{eqn:cox-distribution-final}
G(m) = \sum_{k=0}^m \frac{\Gamma(k+a)}{k!c^k} = \frac{\Gamma(a)}{\left(\frac{c-1}c\right)^a}-\frac{\Gamma(m+1+a)\cdot \hgf{1,m+a+1}{m+2}{\frac 1 c}}{(m+1)!c^{m+1}},
\end{dmath}
where
\begin{equation}\label{eqn:generalized-hypergeometric-function}
    \ghgf{p}{q}{a_1,\ldots,a_p}{b_1,\ldots,b_q}{x} := \sum_{k=0}^\infty
    \frac{(a_1)_k(a_2)_k\cdots(a_p)_k}{(b_1)_k(b_2)_k\cdots(b_q)_k}\frac{x^k}{k!}
\end{equation}
is the generalized hypergeometric function, and
\begin{equation}\label{eqn:pochhammer-symbol}
(x)_n := \frac{\Gamma(x+n)}{\Gamma(x)} = x(x+1)(x+2)\cdots(x+n-1)
\end{equation}
is the Pochhammer symbol. Useful special cases are $(1)_k=k!$, as well as $(2)_k=(k+1)!$.

It is important to notice that the parameters of the Gamma density used for deriving the density
\eqref{eqn:error-probability} satisfy $a,b>0$, in which case $0\leq \frac 1 c=\frac 1{b+1}<1$, and the hypergeometric series
in \eqref{eqn:cox-distribution-final} converges absolutely for this
argument, by D'Lambert's quotient criterion (see \cite{Slater1966}). This is crucial for the permission to rearrange
the infinite sums in later stages of the upcoming derivation.

We verify expression \eqref{eqn:cox-distribution-final} by induction. For $m=0$, the hypergeometric function evaluates
to (using \eqref{eqn:generalized-hypergeometric-function})
\begin{eqnarray*}
    a\left[\hgf{1,1+a}{2}{\frac 1 c}\right] &=& a\sum_{k=0}^\infty \frac{(1)_k(1+a)_k}{(2)_k}\frac 1{k!c^k}\\
    &=& \sum_{k=0}^\infty \frac{(a)_{k+1}}{(k+1)!}\frac 1{c^k}\\
    &=& c\sum_{k=1}^\infty \frac{(a)_k}{k!}\frac 1{c^k}.\\
\end{eqnarray*}
The last series can be written in the form
\begin{equation}\label{eqn:shifted-sum}
\sum_{k=1}^\infty \frac{(a)_k}{k!}\alpha_k\frac 1{c^k} = -1 + \sum_{k=0}^\infty \frac{(a)_k}{k!}\alpha_k\frac 1{c^k}
\end{equation}
with coefficients $\alpha_k = 1$ for all $k$. This permits the application of an identity due to Euler (cf.
\cite{Norlund1955,Weisstein,Weissteina}),
\begin{equation}\label{eqn:euler-identity}
\sum_{k=0}^\infty \frac{(a)_k}{k!}\alpha_kz^k = (1-z)^{-a}\sum_{k=0}^\infty \frac{(a)_k}{k!}\Delta^k\alpha_0\left(\frac z{1-z}\right)^k,
\end{equation}
with the forward difference $\Delta^n \alpha_0$ defined as
\begin{equation}\label{eqn:forward-difference}
    \Delta^n\alpha_0 = \sum_{i=0}^n (-1)^i \binom n i \alpha_{n-i}.
\end{equation}
Expression \eqref{eqn:forward-difference} can be evaluated directly using the Binomial formula:
\[
    \Delta^n a_0 = \sum_{i=0}^n (-1)^i\binom n i 1^{n-i} = \left\{
                                                             \begin{array}{ll}
                                                               (-1+1)^n = 0, & \hbox{for }n>0 \\
                                                               (-1)^0\binom 0 0 1^0= 1, & \hbox{for } n = 0.
                                                             \end{array}
                                                           \right.
\]
Hence, expression \eqref{eqn:shifted-sum}, by setting $z:=\frac 1 c$ and thanks to the identity
\eqref{eqn:euler-identity} becomes (cf. also \cite[pg. 46]{Slater1966})
\[
    \sum_{k=1}^\infty \frac{(a)_k}{k!}\frac 1{c^k}=\left(1-\frac 1 c\right)^{-a}-1,
\]
and furthermore
\begin{equation}\label{eqn:hgf-evaluated}
a\left[\hgf{1,1+a}{2}{\frac 1 c}\right] = c\left(\left(1-\frac 1 c\right)^{-a}-1\right).
\end{equation}
To accomplish the induction start at $m=0$, we need to verify that (cf. Equation \eqref{eqn:cox-distribution-final})
\begin{equation}\label{eqn:cox-dist-ind-start}
\Gamma(a)\stackrel ? = \frac{\Gamma(a)}{\left(\frac{c-1}c\right)^a}-\frac{\hgf{1,a+1}{2}{\frac 1 c}\Gamma(1+a)}c.
\end{equation}
The identity $\Gamma(1+a)=a\Gamma(a)$, in connection with Equation \eqref{eqn:hgf-evaluated} then gives
\[
\frac{\Gamma(a)}{\left(\frac{c-1}c\right)^a} - \frac{\Gamma(a)c\left(\left(1-\frac 1 c\right)^{-a}-1\right)}c = \Gamma(a),
\]
and claim \eqref{eqn:cox-dist-ind-start} is proved.

Now, assume the formula to be valid up to $m-1$. To accomplish the induction step, let us look at the difference
$G(m+1)-G(m)$ (cf. Equation \eqref{eqn:cox-distribution-final}), which we need to prove equal to the $(m+1)$-th term in
the series \eqref{eqn:cox-distribution}. The difference between the $m$-th term and the $(m+1)$-th term of $G(m+1)$
is
\begin{dmath*}
\frac{\Gamma(m+1+a)\cdot \hgf{1,m+1+a}{m+2}{\frac 1 c}}{(m+1)!c^{m+1}}-\frac{\Gamma(m+2+a)\cdot \hgf{1,m+2+a}{m+3}{\frac 1 c}}{(m+2)!c^{m+2}}
\end{dmath*}
Using the common denominator $(m+2)!c^{m+2}$ and the relation $\Gamma(x+1)=x\Gamma(x)$ to get
$\Gamma(m+2+a)=(m+1+a)\Gamma(m+1+a)$, we can substitute
\[
    A :=\hgf{1,m+2+a}{m+3}{\frac 1 c}, B :=\hgf{1,m+1+a}{m+2}{\frac 1 c}
\]
into the last expression, to obtain the following equality, which is to be verified
\[
    \frac{c(m+2)\Gamma(m+1+a)B-(m+1+a)\Gamma(m+1+a)A}{(m+2)!c^{m+2}} \stackrel ? = \frac{\Gamma(m+a+1)}{(m+1)!c^{m+1}},
\]
where the right hand side is the $(m+1)$-th term in the sum \eqref{eqn:cox-distribution}. The second equality follows
from $\Gamma(m)=(m-1)!$, as $m$ is an integer. Canceling the terms $\Gamma(m+1+a), (m+1)!$ and $c^{m+1}$ on both sides
leaves us with
\[
    \frac{c(m+2)B-(m+1+a)A}{(m+2)c}\stackrel ? = 1.
\]
Dividing the nominator and denominator on the left hand side by $c(m+2)$, we need to verify if
\begin{equation}\label{eqn:hgf-difference}
    B -\frac{m+1+a}{(m+2)c}A\stackrel ? = 1.
\end{equation}
Consider only the right term in the difference, and substitute the expression for $A$. Then we find
\begin{eqnarray}
    \frac{m+1+a}{(m+2)c}\sum_{k=0}^\infty\frac{(1)_k(m+2+a)_k}{(m+3)_k}\frac 1 {k!c^k} &=& \sum_{k=0}^\infty\frac{(m+1+a)_{k+1}}{(m+2)_{k+1}}\frac 1{c^{k+1}}\nonumber\\
    &=& \sum_{k=1}^\infty\frac{(m+1+a)_k}{(m+2)_k}\frac 1{c^k}\label{eqn:hgf-difference-lhs}
\end{eqnarray}
by definition of the Pochhammer symbol \eqref{eqn:pochhammer-symbol}. For the other term in \eqref{eqn:hgf-difference},
we find by definition of the hypergeometric function \eqref{eqn:generalized-hypergeometric-function},
\begin{equation}\label{eqn:hgf-difference-rhs}
    B = \sum_{k=0}^\infty \frac{(1)_k(m+1+a)_k}{(m+2)_k}\frac 1{k!c^k}=\sum_{k=0}^\infty \frac{(m+1+a)_k}{(m+2)_k}\frac 1{c^k}
\end{equation}
Subtracting expression \eqref{eqn:hgf-difference-lhs} from expression \eqref{eqn:hgf-difference-rhs} leaves only the
$0$-th term in the sum, which is equal to 1, and \eqref{eqn:hgf-difference} is proved.

The distribution function in its complete form is finally obtained by substituting $c=b+1$, and plugging
\eqref{eqn:cox-distribution-final} into \eqref{eqn:cox-distribution}, giving
\begin{dmath*}
    F(m) = {\Prob{X\leq m}}
=\frac{b^a}{(b+1)^a\Gamma(a)}\left[\frac{\Gamma(a)}{\left(\frac b{b+1}\right)^a}-\frac{\Gamma(m+1+a)\cdot \hgf{1,m+a+1}{m+2}{\frac 1 {b+1}}}{(m+1)!(b+1)^{m+1}}\right]
= 1 - \frac{b^a}{(b+1)^a\Gamma(a)}\frac{\Gamma(m+1+a)\cdot \hgf{1,m+a+1}{m+2}{\frac 1 {b+1}}}{(m+1)!(b+1)^{m+1}}
= 1 - \frac{b^a(a)_{m+1}}{(m+1)!(b+1)^{a+m+1}}\hgf{1,m+a+1}{m+2}{\frac 1 {b+1}}.
\end{dmath*}

\subsubsection*{Expected Number of Errors}

Since our model assumes Poissonian error scattering with a gamma-distributed intensity-parameter, recall that if the random
variable $X$ is Poissonian with parameter $\lambda$, then $\E X=\lambda$. But $\lambda$ is gamma-distributed with
parameters $a,b>0$, so the average number of errors per time unit will come to the average error rate, which in turn
is the first moment of the Gamma-distribution, and hence found as
\[
    \E X = \frac a b.
\]
Alternatively, one can verify the above relation by carrying out similar calculations as for obtaining the
distributions function. The only additional task is then a limit process, which can be tackled in a very similar way as
shown below.

\subsubsection*{Parity Check Failure}

We wish to prove that the probability for an odd number of errors is given by \eqref{eqn:parity-miss}, and that for a given string that is $2m+1$ bit long, the probability of having an odd number of errors is \eqref{eqn:finite-parity-miss}.

We prove \eqref{eqn:parity-miss} by first proving \eqref{eqn:finite-parity-miss} by induction, and
then letting $m$ approach infinity. Using \eqref{eqn:cox-distribution}, the probability of an odd number of errors
in a string of length $2m+1$ is proportional to
\begin{dmath}\label{eqn:odd-error-prob}
    \sum_{k=0}^m \frac{\Gamma(2k+1+a)}{(2k+1)!c^{2k+1}}=\frac{\Gamma(a+1)\left(\left(1+\frac 1 c\right)^a-\left(1-\frac 1 c\right)^a\right)}{2a\left(1+\frac 1 c\right)^a\left(1-\frac 1 c\right)^a}
-\frac{\Gamma(2m+3+a)\cdot\ghgf 3 2{1,m+2+\frac a 2,m+\frac 3 2+\frac a 2}{m+2,m+\frac 5 2}{\frac 1{c^2}}}{(2m+3)!c^{2m+3}},
\end{dmath}
where the equality can be obtained using the \textsc{Maple} software package. We verify this equality by induction.
Equations \eqref{eqn:parity-miss} and \eqref{eqn:finite-parity-miss} are obtained by substituting
$c=b+1$, multiplying with $\frac{b^a}{(b+1)^a\Gamma(a)}$, taking the limit $m\to\infty$ and re-arranging terms.

\underline{Induction start}: Substitute $m=0$ into the last expression, then the problem is to verify whether
\begin{dmath*}
    \frac{\Gamma(a+1)\left(\left(1+\frac 1 c\right)^a-\left(1-\frac 1 c\right)^a\right)}{2a\left(1+\frac 1 c\right)^a\left(1-\frac 1 c\right)^a}-\frac{\Gamma(a+3)\cdot\hgf{1,2+\frac a 2,\frac 3 2+\frac a 2}{2,\frac 5 2}{\frac 1 {c^2}}}{6c^3}
\stackrel ? = \frac{\Gamma(a+1)}c
\end{dmath*}
is true. By applying the identity $\Gamma(a+3)=(a+1)(a+2)\Gamma(a+1)$ and multiplying with $c$, we can
cancel $\Gamma(a+1)$ in each term to get
\begin{dmath}\label{eqn:oddness-induction-start}
    \left[\frac c 2 \frac{\left(\left(1+\frac 1 c\right)^a-\left(1-\frac 1 c\right)^a\right)}{a\left(1+\frac 1 c\right)^a\left(1-\frac 1 c\right)^a}-1\right]-\frac{(a+1)(a+2)\cdot\hgf{1,2+\frac a 2,\frac 3 2+\frac a 2}{2,\frac 5 2}{\frac 1 {c^2}}}{6c^2}\stackrel ? = 0 .
\end{dmath}
The hypergeometric function is by definition
\begin{eqnarray}
\hgf{1,2+\frac a 2,\frac 3 2+\frac a 2}{2,\frac 5 2}{\frac 1 {c^2}}&=&\sum_{k=0}^\infty \frac{(1)_k \left(2+\frac a 2\right)_k\left(\frac 3 2 +\frac a 2\right)_k}{(2)_k\left(\frac 5 2\right)_k}\frac 1 {k!c^{2k}}\nonumber\\
&=&\sum_{k=0}^\infty \frac{\left(2+\frac a 2\right)_k\left(\frac 3 2 +\frac a 2\right)_k}{(k+1)!\left(\frac 5 2\right)_k}\frac 1 {c^{2k}}.\label{eqn:hgf-eval-2}
\end{eqnarray}
The Pochhammer symbol satisfies the following identities, which we can use to simplify the terms in the series (cf.
\cite{Weissteina}):
\begin{eqnarray}
  (a)_k\left(a+\frac 1 2\right)_k &=& \frac 1{4^k}(2a)_{2k}, \label{eqn:pochhammer-1}\\
  (a+1)_k &=& \frac{a+k}a(a)_k, \label{eqn:pochhammer-2}\\
  \left(\frac 3 2\right)_k &=& \frac{(2k+1)!}{k!4^k}.\label{eqn:pochhammer-3}
\end{eqnarray}
The term $\left(\frac 5 2\right)_k=\left(\frac 3 2+1\right)_k$ can be evaluated using \eqref{eqn:pochhammer-2} and \eqref{eqn:pochhammer-3} to give
\begin{equation}\label{eqn:hgf-denominator}
    \left(\frac 5 2\right)_k = \frac{3+2k}3\cdot\frac{(2k+1)!}{k!4^k}.
\end{equation}
The nominator of the terms in the series \eqref{eqn:hgf-eval-2} are found using \eqref{eqn:pochhammer-1} as
\begin{equation}\label{eqn:hgf-nominator}
\left(\frac 3 2 +\frac a 2\right)_k\left(2+\frac a 2\right)_k = \frac 1 {4^k}\left(2\left(\frac 3 2+\frac a 2\right)\right)_{2k}=\frac 1 {4^k}(a+3)_{2k}.
\end{equation}
Substituting \eqref{eqn:hgf-denominator} and \eqref{eqn:hgf-nominator} into \eqref{eqn:hgf-eval-2} gives
\begin{dmath*}
\hgf{1,2+\frac a 2,\frac 3 2+\frac a 2}{2,\frac 5 2}{\frac 1 {c^2}}(a+1)(a+2)=(a+1)(a+2)\sum_{k=0}^\infty \frac{\frac 1{4^k}(a+3)_{2k}}{(k+1)!\frac{2k+3}3\frac{(2k+1)!}{k!4^k}}\frac 1 {c^{2k}}
\qquad\qquad =3\sum_{k=0}^\infty \frac{(a+1)_{2k+2}}{(k+1)(2k+3)(2k+1)!}\frac 1{c^{2k}}=3c^2\sum_{k=1}^\infty\frac{(a+1)_{2k}}{k(2k+1)(2k-1)!}\frac 1{c^{2k}}
\qquad\qquad =3c^2\sum_{k=1}^\infty\frac{2k(a+1)_{2k}}{k(2k+1)!}\frac 1{c^{2k}}=6c^2\sum_{k=1}^\infty\frac{(a+1)_{2k}}{(2k+1)!}\frac 1{c^{2k}},
\end{dmath*}

so that
\[
 \frac{\Gamma(a+3)\hgf{1,2+\frac a 2,\frac 3 2+\frac a 2}{2,\frac 5 2}{\frac 1 {c^2}}}{6c^2} =
\sum_{k=1}^\infty\frac{(a+1)_{2k}}{(2k+1)!}\frac 1{c^{2k}}.
\]
Substituting $z:=\frac 1 c$ in the last expression, as well as in the term in square brackets in
\eqref{eqn:oddness-induction-start}, our task is to verify whether
\[
\sum_{k=1}^\infty\frac{(a+1)_{2k}}{(2k+1)!}z^{2k} \stackrel ? = \frac 1 {2az}\left[\frac 1{(1-z)^a}-\frac 1{(1+z)^a}\right]-1
\]
holds. This is achieved, by forming the Taylor-series expansion of the right hand side around $z=0$, turning out equal
to the series on the left side.

\underline{Induction step}: To accomplish the induction step, we compare the difference between the results when
substituting $m+1$ and $m$ into \eqref{eqn:odd-error-prob}, which should be equal to the $(m+1)$-th term in the finite
sum.

This difference comes to
\begin{dmath*}
    \frac{\Gamma(2m+3+a)}{(2m+3)!c^{2m+3}}  \stackrel ? =\frac{\Gamma(2m+3+a)\cdot\ghgf 3 2{1,m+2+\frac a 2,m+\frac 3 2 +\frac a 2}{m+2,m+\frac 5 2}{\frac 1{c^2}}}{(2m+3)!c^{2m+3}}
     \qquad -\frac{\Gamma(2m+5+a)\cdot\ghgf 3 2{1,m+3+\frac a 2,m+\frac 5 2 +\frac a 2}{m+3,m+\frac 7 2}{\frac 1{c^2}}}{(2m+5)!c^{2m+5}}.
\end{dmath*}

Using the identity $\Gamma(2m+5+a)=(2m+3+a)(2m+4+a)\Gamma(2m+3+a)$, and dividing by the right hand side leaves us with the claim
\begin{dmath*}
    1  \stackrel ? = \ghgf 3 2{1,m+2+\frac a 2,m+\frac 3 2 +\frac a 2}{m+2,m+\frac 5 2}{\frac 1{c^2}}
    \qquad -\left[\frac{(2m+3+a)(2m+4+a)}{(2m+4)(2m+5)c^2}\cdot\ghgf 3 2{1,m+3+\frac a 2,m+\frac 5 2 +\frac a 2}{m+3,m+\frac 7 2}{\frac 1{c^2}}\right]     .
\end{dmath*}
Let us pay closer attention to the term in square brackets. By canceling 2 from all four brackets in the fraction in
front of the hypergeometric function, and writing down the latter as a series, we obtain
\[
    \frac{\left(m+2+\frac a 2\right)\left(m+\frac 3 2 +\frac a 2\right)}{(m+2)\left(m+\frac 5 2\right)c^2}\sum_{k=0}^\infty\frac{\left(m+3+\frac a 2\right)_k\left(m+\frac 5 2+\frac a 2\right)_k}{\left(m+3\right)_k\left(m+\frac 7 2\right)_k}\frac 1{c^{2k}}
\]
Using the identity $a(a+1)_n = (a)_{n+1}$, we can assemble the nominator and the denominator into the sum's terms to
find
\begin{eqnarray*}
&& \frac{\left(m+2+\frac a 2\right)\left(m+\frac 3 2 +\frac a 2\right)}{(m+2)\left(m+\frac 5 2\right)c^2}\sum_{k=0}^\infty\frac{\left(m+3+\frac a 2\right)_k\left(m+\frac 5 2+\frac a 2\right)_k}{(m+3)_k\left(m+\frac 7 2\right)_k}\frac 1{c^{2k}}\\
&& \qquad\qquad = \sum_{k=0}^\infty\frac{\left(m+2+\frac a 2\right)_{k+1}\left(m+\frac 3 2+\frac a 2\right)_{k+1}}{(m+2)_{k+1}\left(m+\frac 5 2\right)_{k+1}}\frac 1{c^{2(k+1)}}\\
&& \qquad\qquad = \sum_{k=1}^\infty\frac{\left(m+2+\frac a 2\right)_k\left(m+\frac 3 2+\frac a 2\right)_k}{(m+2)_k\left(m+\frac 5 2\right)_k}\frac 1{c^{2k}}\\
&& \qquad\qquad = \ghgf 3 2{1,m+2+\frac a 2,m+\frac 3 2 +\frac a 2}{m+2,m+\frac 5 2}{\frac 1{c^2}}-1
\end{eqnarray*}
and the claim is proved.

\subsubsection*{Limit for $m\To\infty$}

Our final task is calculating
\[
    \lim_{m\To\infty}\frac{\Gamma(2m+3+a)}{(2m+3)!c^{2m+3}}\cdot\ghgf 3 2{1,m+2+\frac a 2,m+\frac 3 2+\frac a 2}{m+2,m+\frac 5 2}{\frac 1{c^2}}.
\]

To get rid of the generalized hypergeometric function, let us upper-bound the
series by upper-bounding each term separately. We have
\begin{dmath*}
    \ghgf 3 2{1,m+2+\frac a 2,m+\frac 3 2+\frac a 2}{m+2,m+\frac 5 2}{\frac 1{c^2}}=\sum_{k=0}^\infty\frac{\left(m+2+\frac a 2\right)_k\left(m+\frac 3 2+\frac a 2\right)_k}{(m+2)_k\left(m+\frac 5 2\right)_k}\frac 1{c^{2k}}
 \leq  \sum_{k=0}^\infty\frac{[(m+2+a)_k]^2}{[(m+2)_k]^2}\frac 1{c^{2k}}
 \leq  \left[\sum_{k=0}^\infty \frac{(m+2+a)_k}{(m+2)_k}\frac 1 {c^k}\right]^2,
\end{dmath*}
where we have used the inequality $(x)_k\leq (y)_k$, for $0\leq x\leq y$. Because $m$ is an integer, we can write $(2m+3)!=\Gamma(2m+4)$. By
substituting $z:=\frac 1 c>0$ we obtain
\begin{equation}\label{eqn:bounding-series}
\frac{\Gamma(2m+3+a)}{(2m+3)!}\cdot{_3F_2}(\ldots)\leq \left[\sum_{k=0}^\infty \underbrace{\frac{\Gamma(2m+3+a)^2}{\Gamma(2m+4)^2}\frac{(m+2+a)_k}{(m+2)_k}}_{=:\alpha(k)}z^k\right]^2.
\end{equation}
Expanding the Pochhammer symbols on the right hand side in terms of the gamma function, the coefficient $\alpha(k)$ of
$z^k$ becomes
\[
\alpha(k)=\underbrace{\Gamma(2m+3+a)^2}_{=:[P(m)]^2}\underbrace{\frac 1{\Gamma(2m+4)^2}}_{=:[Q(m)]^2}\underbrace{\frac{\Gamma(m+3+a+k)}{\Gamma(m+2+k)}}_{=:R(m)}\underbrace{\frac{\Gamma(m+2)}{\Gamma(m+3+a)}}_{=:S(m)}
\]
Considering the terms $P(m),Q(m),R(m),S(m)$ separately significantly simplifies matters, when we apply D'Alambert's
quotient-criterion to investigate the convergence of the series \eqref{eqn:bounding-series}. The quotient of interest
is
\[
\abs{\frac{\alpha_{k+1}z^{k+1}}{\alpha_kz^k}} = z\frac{P(m+1)^2}{P(m)^2}\frac{Q(m+1)^2}{Q(m)^2}\frac{R(m+1)}{R(m)}\frac{S(m+1)}{S(m)},
\]
where positive values are guaranteed since all involved quantities are positive. Let us consider the four quotients
individually:
\begin{enumerate}
  \item[i)] $P(m+1)/P(m)$: This is
\begin{equation}\label{eqn:p-quotient}
    \frac{P(m+1)}{P(m)} = \frac{\Gamma(2m+5+a)}{\Gamma(2m+3+a)} = (2m+3+a)(2m+4+a),
\end{equation}
because $\Gamma(2m+5+a)=(2m+3+a)(2m+4+a)\Gamma(2m+3+a)$.
  \item[ii)] $Q(m+1)/Q(m)$: Using the same reasoning as before, we get
    \begin{equation}\label{eqn:q-quotient}
        \frac{Q(m+1)}{Q(m)} = \frac{\Gamma(2m+4)}{\Gamma(2m+6)}=\frac 1 {(2m+4)(2m+5)}.
    \end{equation}
    Multiplying \eqref{eqn:p-quotient} with \eqref{eqn:q-quotient}, we obtain a rational function with polynomials
    of equal order and leading coefficient in the nominator and denominator. It follows that
    \[
        \lim_{m\To\infty}\frac{P(m+1)^2}{P(m)^2}\frac{Q(m+1)^2}{Q(m)^2} = 1
    \]
  \item[iii)] $S(m+1)/S(m)$: Once more, exploiting the recurrence relation for the $\Gamma$-function, $\Gamma(m+3)=\Gamma(m+2)(m+2)$ and $\Gamma(m+4+a)=(m+3+a)\Gamma(m+3+a)$, we find
    \[
        \frac{S(m+1)}{S(m)} = \frac{m+2}{m+3+a}\To 1,\quad\text{as }m\To\infty.
    \]
\item[iv)] $R(m+1)/R(m)$: Analogously as for the quotient $S(m+1)/S(m)$,
\[
    \frac{R(m+1)}{R(m)}=\frac{m+3+a+k}{m+2+k} = 1+\frac 1{m+2+k} \leq 1+\frac 1{m+2},
\] 
where the second inequality is valid since $k\geq 1$.
\end{enumerate}

Now, choose two constants $\rho_1,\rho_2>1$ such that $z\rho_1\rho_2<1$, which is possible since $0<z<1$ is itself a constant (recall that $z=\frac 1 c=\frac 1{b+1}$ and $b>0$). Convergence of all quotients implies the existence of constants $M_1, M_2$ such that
$[P(m+1)Q(m+1)]^2/[P(m)Q(m)]^2\leq\rho_1$ for $m\geq M_1$. Furthermore, we have $S(m+1)/S(m)\leq 1$ for all $m\geq 0$ (because $a>0$), and
finally, $R(m+1)/R(m)\leq\rho_2$ for every $m\geq M_2$. Setting $q:=z\rho_1\rho_2$ and choosing $M:=\max\set{M_1,M_2}$,
we find a uniform bound for \eqref{eqn:bounding-series} given by the geometric series with quotient $q<1$. Hence, the
series \eqref{eqn:bounding-series} is ultimately bounded by a constant $L>0$ for every $m>M$. So
\[
\frac{\Gamma(2m+3+a)}{(2m+3)!}\cdot{_3F_2}(\ldots)\leq L^2
\]
for $m>M$, and furthermore,
\begin{dmath*}
0\leq\lim_{m\To\infty}\frac{\Gamma(2m+3+a)}{(2m+3)!c^{2m+3}}\ghgf 3 2{1,m+2+\frac a 2,m+\frac 3 2+\frac a 2}{m+2,m+\frac 5 2}{\frac 1{c^2}}\leq{\lim_{m\To\infty}\frac{L^2}{c^{2m+3}} = 0},
\end{dmath*}
since $c=b+1>1$.


\end{document}